# High-Entropy Oxides Based on Valence Combinations: Design and Practice

*Lei Tang, Zemin Li, Kepi Chen*, Cuiwei Li, Xiaowen Zhang, and Linan An**

L. Tang, Z. Li, Prof. K. Chen
School of Energy, Power and Mechanical Engineering
North China Electric Power University
Beijing 102206, China
E-mail: ckp@ncepu.edu.cn

Prof. C. Li
School of Mechanical, Electronic and Control Engineering
Beijing Jiaotong University
Beijing 100044, China

Prof. X. Zhang
State Key Laboratory of New Ceramics and Fine Processing
School of Materials Science and Engineering
Tsinghua University
Beijing 100084, China

Prof. L. An
Department of Materials Science and Engineering
University of Central Florida
Orlando, FL 32816, USA
E-mail: linan.an@ucf.edu

**Abstract:**

High-entropy oxides (HEOs) are a new class of materials that are promising for a wide range of applications. Designing HEOs needs to consider both geometric compatibility and electrical equilibrium. However, there is currently no available method to systematically consider these two factors when selecting constituent materials for making HEOs. Here we propose a two-step strategy, where a HEO system to be explored is first partitioned into multiple subsystems based on the valence combinations of substituted cations; the geometric compatibility is then considered in selecting suitable substituted cations. We demonstrate this strategy by using $A(5B_{0.2})O_3$ perovskite as a model system. We show that the system can be partitioned into 12 subsystems. Ten of the subsystems have formed a single-phase cubic perovskite, while two have partially ordering structure. The formation of single phases is correlated to Goldschmidt's tolerance factor, while the formation of the ordering structure is mainly correlated to cation-valence difference. We anticipate that this strategy is applicable to exploring HEOs in other systems.

The rapid evolution in modern technology has put forward a pressing demand for developing new materials with advanced properties. Recently, this demand has led to the discovery of a new family of materials named high-entropy alloys (HEAs), where five or more elemental species are deliberately incorporated into a single lattice with random occupancy[1-3]. These multicomponent materials exhibited many extremely attractive properties, surpassing those of the constituent materials[4-6]. Inspired by the prosperity of HEAs, the first high-entropy oxide, entropy-stabilized (MgCoNiCuZn)O, was reported in 2015[7]. Thereafter, HEOs with various crystalline structures[8-13], as well as other high-entropy ceramics (HECs)[14-18], have been demonstrated. Although still in their infancy, HECs have already shown promising properties for a wide range of applications including thermoelectricity[19], thermal and environmental protection[20, 21], energy storage[22-26], and water splitting[27]. It is believed that the remarkable properties of these high-entropy materials (HEMs) originated from their high disorder and lattice distortion[2, 3, 28, 29].

Selecting substituted cations for HEOs requires two factors to be considered. The first factor is geometric compatibility, which requires that the size of each substituted cation should be appropriate for the corresponding oxygen polyhedron and the average size of these cations should accord to structural tolerance factors (e.g. Goldschmid structural tolerance factor)[30]. The second factor is electrical equilibrium, which requires that substituted cations must have a special combination of valences so that their average valence is equal to the value required to maintain stoichiometry (although slight non-stoichiometry is allowed). So far, this last requirement has received little attention since previously reported HEOs were primarily made from substituted cations with the same valence (or were non-stoichiometry) [9-12]. Using substituted cations with different valences to design HEOs have not been deliberately explored. Such negligence is surprising because the use of substituted cations with different valences can greatly expand the range of cation choices, and thus the HEO family. In addition, since the electrostatic bond strength to oxygen for a given polyhedron is determined by the valence of the cation, the valence mismatch between the cations in the two polyhedra sharing the oxygen will cause distortion of the electron cloud distribution (referred to herein as charge distortion). Consequently, the use of cations with different valences to make HEOs gives an additional dimension to distort the materials, which is unique to HEOs.

A strategy is highly desired when exploring systems with more than one variable and the number of samples in the system is extremely large, such as HEOs. A good strategy should ensure that the entire system to be explored will be covered and repeat exploration will be avoided. Here, we propose the following two-step strategy for better exploring HEOs. First, we determine the possible independent valance combinations of substituted cation by considering electrical equilibrium and partition the system into multiple subsystems according to these combinations. That is, materials with the same combination of substituted cation valence will be grouped into one subsystem. Second, we select suitable cations for each subsystem by considering geometric compatibility. The use of valence combination to partition HEOs is due to the fact that the valence of the substituted cations is an integer within a certain range, so that for a given system, the number of independent valence combinations (or subsystems) is finite, and the demarcation between these combinations (subsystems) is unambiguous.

In this study, we use $ABO_3$ type high-entropy perovskite as a model system to demonstrate the proposed strategy. Specifically, we focus on the system where the A-site was occupied by $Ba^{2+}$ and the B-site was occupied by five substituted cations with different valences in equiatomic amount (referred to as $Ba(5B_{0.2})O_3$). Perovskite oxides are one of the most important functional materials with a multiscale controllable structure and diverse properties. The widespread applications of the materials include dielectrics, ferroelectric and multiferroic devices, cathode for solid oxide fuel cells, piezoelectric devices, superconductors, and photovoltaic modules. A unique feature of perovskite oxides is that both their A-site and B-site can be substituted with a wide range of cations to selectively tailor the multiple physical properties of the materials. This ability can be greatly expanded by forming high-entropy compounds. High-entropy perovskite oxides (HEPOs) with either A-site, B-site or both being substituted with 5 different cations in equiatomic amounts have been reported recently[11, 12, 31, 32]. As we know, HEPOs reported so far were primarily made of the substituted cations with the same valence. Thus, the current study also represents the first report of high-entropy oxides made of substituted cations with different valences.

First, we partition $Ba(5B_{0.2})O_3$ system into subsystems based on the valence combination of substituted cations. After considering various factors (such as the availability and stability of raw materials), we limit the valences of the substituted cations to +2 to +6. If there is no other restriction, the number of possible valence combinations of the five substituted cations is $5^5 = 3125$. Electrical equilibrium requires that in order to form stoichiometric $ABO_3$ compounds, the average valence of the five B-site substituted cations must be +4. This restriction reduces the number of independent valence combinations to 12. That is to say, the $Ba(5B_{0.2})O_3$ system has 12 subsystems. These 12 subsystems (independent valence combinations) are summarized in Table 1. It can be seen that only subsystem C12 has been explored[11]. The other eleven subsystems have not yet been deliberately explored. The subsystems are also determined for $ABO_3$ systems containing 4 or 6 B-site substituted cations in equiatomic amount (Supporting Information, Table S1 and S2).

Next, we select cations suitable for synthesizing high-entropy $Ba(5B_{0.2})O_3$ in each subsystem by considering geometric compatibility. Note that although each subsystem may contain multiple materials, the purpose of this study is by no means to explore all possible materials, but to use one material in each subsystem to demonstrate the proposed strategy. Two geometric restrictions need to be considered when selecting cations to form a B-site substituted perovskite. First, the size of the cations should be suitable for the oxygen polyhedron, which is an octahedron for the B-site of perovskite. According to Pauling's first rule[33], the ratio of the radius of substituted cation to the radius of oxygen ion should be between 0.414 to 0.732. That is, the radius of the substituted cations should be between 0.58 Å to 1.025 Å. Second, the average radius of the selected cations should meet the requirement of the structural tolerance factor, *t*, proposed by Goldschmid[30]

$$t = \frac{r_A + r_O}{\sqrt{2}(r_B + r_O)} \qquad (1)$$

where $r_A$, $r_B$, and $r_O$ are the ionic radii of the cation at A-site, the cation at B-site and oxygen ion, respectively. A recent study has revealed that in order to form stable cubic perovskite, the value of *t* should be around 1[11]. The five cations selected for each subsystem are listed in Table 1. The coordination numbers and radii of the ions used in this study are given in Table 3 of Supporting Information.

We then evaluate whether these compositions can form single-phase high-entropy perovskite. To do so, pellets made of these compositions were sintered at various temperatures. The resultant samples were characterized by using XRD and SEM-EDXS (Figs. 1 and 2, and Figs. S1-S10 in Supporting Information). Key findings are summarized in Table 2. It was found that ten out of the twelve compositions can form a single-phase cubic perovskite within certain temperature ranges. Fig. 1a shows the XRD patterns of the samples made from composition C12, indicating that the composition can form a single-phase perovskite in a temperature range of 1300 °C – 1400 °C without a detectable second phase. Fig. 1b shows the typical EDXS elemental mappings of the sample made at 1400 $^oC$. No element-enriched secondary phase can be identified within the resolution, confirming the formation of a single phase. The XRD pattern was also calculated for each composition using the rule of mixture by assuming equal and random distribution of five substituted cations. For the compositions that can form a single phase, the calculated XRD patterns agree well with the experimentally measured XRD patterns, indicating that substituted cations likely randomly occupied the B-site. A previous study suggested that formation of a single-phase cubic perovskite is correlated with the structural tolerance factor[11]. The factor was calculated for each composition studied here using the average radius of the substituted cations for $r_B$ which are listed in Table 1. The value of $t$ varies between 0.9 to 1.04. It is seen that with the exception of composition C5, the compositions with $0.97 \leq t \leq 1.03$ can form a single cubic perovskite phase, while the composition (C1) with $t \geq 1.03$ cannot form a single phase. This result agrees well with the previous study[11].

Interestingly, the XRD patterns suggest that the samples made from 2 compositions (C2 and C3) contain the (111) chemical ordering of the B-site. Fig. 2 shows the XRD patterns of the samples made from composition C2. A diffraction peaks corresponding to the ordering can be identified from the samples made at both 1200 $^oC$ and 1350 $^oC$. Such an ordering structure has not been observed in high-entropy oxides before. Previous studies revealed that the formation of the ordering structure in the sample depended on ionic size and valence differences of the substituted cations[34,35]. In order to assess the influence of cationic size on the formation of the order structure, we calculated the B-site cation-size difference using the following equation,

$$\delta(r_B) = \sqrt{\sum_{i=1}^{5} c_i \left(1 - \frac{r_i}{\sum_{i=1}^{5} c_i r_i}\right)^2} \qquad (2)$$

where $r_i$ and $c_i$ are the radius and the mole fraction of the $i^{th}$ cation, respectively. The result (Table 1) suggests that for the materials studied here, there is no correlation between the formation of the ordered structures and the cation-size difference. In order to assess whether the formation of the order structures is due to valence mismatch, we introduce a parameter to quantitatively describe the valence mismatch, named B-site cation-valence difference ($\delta(V_B)$) :

$$\delta(V_B) = \sqrt{\sum_{i=1}^{5} c_i (1 - \frac{V_i}{\sum_{i=1}^{5} c_i V_i})^2} \quad (3)$$

where $V_i$ is the valence of the $i^{th}$ cation. $\delta(V_B)$ is calculated for the 12 compositions and listed in Table 1. We find that the formation of ordered structures is well correlated with $\delta(V_B)$, where order structure can be formed for the compositions with $\delta(V_B) \geq 0.42$, but cannot be formed for the compositions with $\delta(V_B) \leq 0.35$. This result indicates that the formation of order structure is more correlated with cation-valence difference, instead of cation-size difference, consistent with previous studies[36].

Finally, the room temperature dielectric constant and dielectric loss were measured for the samples made from the 10 compositions that can form single phases (Table 3). The sintering temperature for each composition was selected so that high density can be achieved while the sample maintains a single phase or contains only a small amount of second phase. It can be seen that the dielectric properties of these materials are spread over a wide range. The dielectric constant of the samples varies between 25-526, and the dielectric loss varies from 0.003 to 2.63. This provides an opportunity to choose materials for afferent applications. For example, the sample made from composition C10 having a dielectric constant of 46.5 and dielectric loss of 0.26% could be useful as microwave dielectric materials like $Ba(Mg_{1/3}Nb_{1/2})O_3$. The sample made from C6 exhibits a dielectric constant of 526 and a dielectric loss of 2.63, which could be useful in microwave absorption.

The temperature-dependent alternating current (AC) conductivity of the 10 samples listed in Table 3 was measured in a temperature range of 437 °C – 666 °C at 1 kHz. The results were analyzed using the Arrhenius equation

$$\sigma_{ac} = \sigma_0 \exp(-\frac{E_a}{k_B T}) \qquad (4)$$

Where $\sigma_0$ is a constant, $E_a$ is the activation energy for conduction, $k_B$ is the Boltzmann constant and $T$ is the absolute temperature. As shown in Fig. 3, samples made of C2, C3, C4, C9 and C12 follow the Arrhenius relationship very well over the entire test temperature range, while the rest of samples obeys the Arrhenius relationship only at high temperatures. The activation energy, $E_a$, is calculated for each sample and is listed in Table 3. It is seen that 8 samples have the activation energy in the range of 0.98eV ~ 1.21eV. These values are very close to half of the intrinsic band gap width (2.5 ~ 3.2eV) of $BaTiO_3$[37], indicating that these high-entropy perovskites likely have the band structure very similar to that of $BaTiO_3$, which is an intrinsic semiconductor in the test temperature range. The activation energy of the other two samples is 0.7eV, which is smaller than that of barium titanate and strontium titanate ceramics[37,38], suggesting these materials have potential as thermoelectric materials.

In summary, we have proposed a systematic strategy for designing high-entropy oxides, in which a system to be explored is first divided into multiple subsystems by considering electrical balance. The substituted cations are then selected by considering geometrical compatibility. We applied the strategy to the $Ba(5B_{0.2})O_3$ perovskite system, and identified 12 subsystems. Suitable substituted cations were then determined for each subsystem by considering both Pauling's first rule and Goldschmidt's tolerance factor. We found that 10 out of the 12 compositions formed single phase cubic perovskite in the corresponding temperature range, and 2 have (111) chemical ordering structure. The formation of single phase is influenced by the tolerance factor, while the formation of partial ordering structure is mainly influenced by cation-valence difference. The current results demonstrated that the strategy is effective in designing high-entropy oxides, which can be expanded to other oxide systems.

**Experimental Section**

The starting materials used in this study are $BaCO_3$ (99%) powder purchased from Sinopharm Chemical Reagent Co., Ltd., and MgO (99.99%), ZnO (99%), $Y_2O_3$ (99.5%), $Yb_2O_3$ (99.99%), $CeO_2$ (99.5%), $HfO_2$ (98%), $ZrO_2$ (99.99%, 50nm), $SnO_2$ (99.5%), $GeO_2$ (99.99%), $TiO_2$ (99.99%, anatase), $Sb_2O_5$ (99%), $Ta_2O_5$ (99.99%), $Nb_2O_5$ (99.9%), $Ga_2O_3$ (99.99%), $WO_3$ (99.99%) and $MoO_3$ (99.99%) powders purchased from Aladdin Reagent (Shanghai) Co., Ltd.. The as-received powders were first baked at 120 $^oC$ for 24 hr to remove moisture. The dried powders, in an atomic ratio required to form the stoichiometric compounds listed in Table 1, were then uniformly mixed together by ball milling for 24 hr. The powder mixture was calcined at 1200 $^oC$ or 1300 $^oC$ for 12hr. The resultant powder was compressed into discs with a diameter of 15 mm and a thickness of 2 mm, using a small amount of PVB as a binder. The discs were first heat-treated at 550 $^oC$ for 4 hr to burn out PVB, and then sintered at 1200 $^oC$ to 1600 $^oC$ for 6 hr.

The phase composition of the resultant ceramics was characterized using X-ray diffraction (XRD) with Cu Kα (λ=1.54Å) as the radiation. The microstructure and element distribution were analyzed using field-emission scanning electron microscopy (SEM) with energy dispersive X-ray spectroscopy (EDXS, FESEM, Quanta FEG 650, FEI).

For electric property measurement, silver paste was painted on the surfaces of the discs as the electrodes. The spectra were acquired using a precision LCR meter (E4980AL, KEYSIGHT) in a frequency range of 20 Hz - 1 MHz and a temperature range of 437 $^oC$ - 666 $^oC$.

**Supporting Information**
Supporting Information is available from the Wiley Online Library or from the author.

**Acknowledgements**
L. Tang and Z. Li contributed equally to this work. This work was supported by the National Natural Science Foundation of China (NSFC) under the No. 21371056.

Received: ((will be filled in by the editorial staff))
Revised: ((will be filled in by the editorial staff))
Published online: ((will be filled in by the editorial staff))

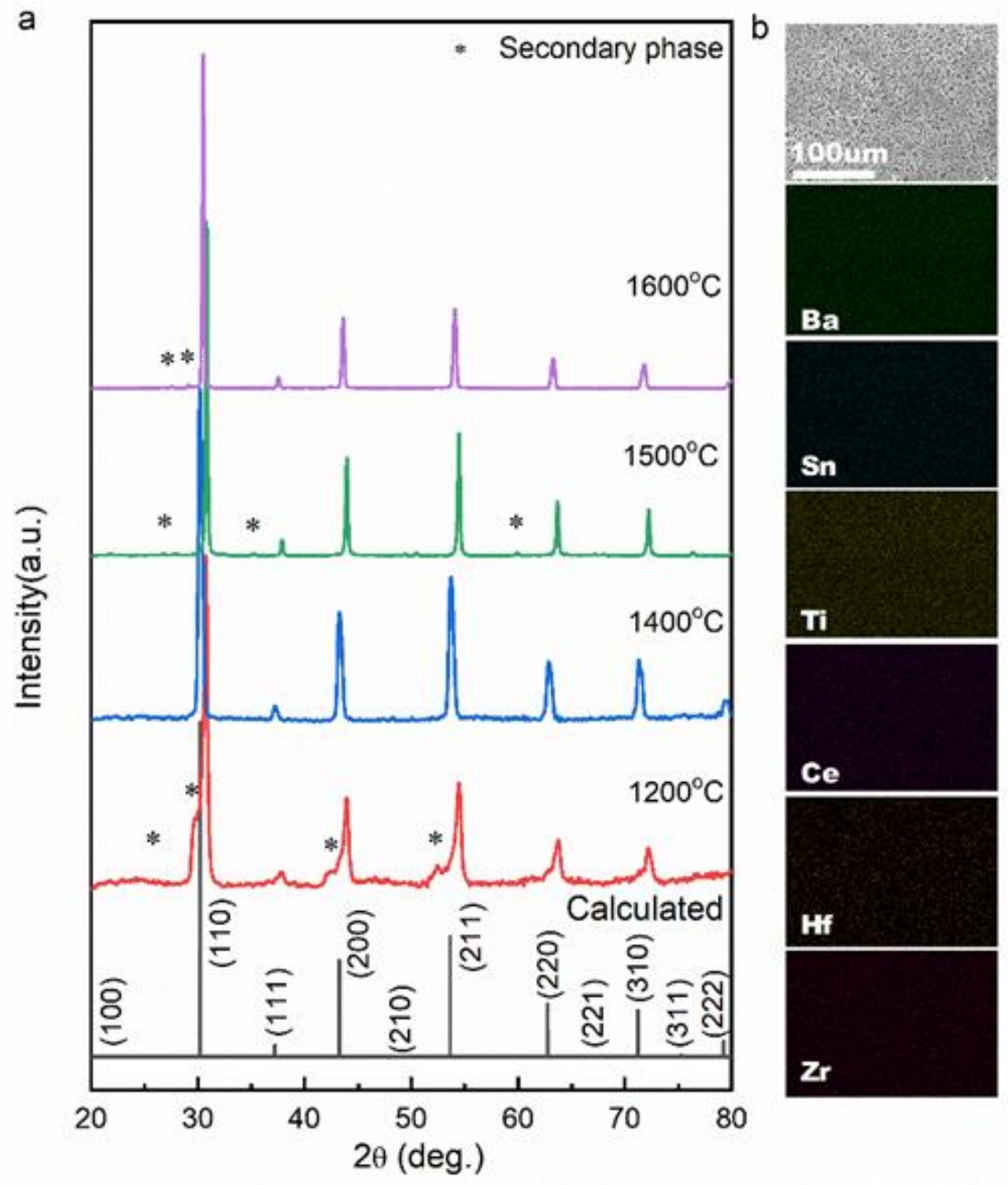

**Figure 1.** (a) XRD patterns of the sample made from composition C12 at various temperatures as labeled. (b) SEM-EDXS elemental maps of the sample made from composition C12 at 1400 $^{o}$C.

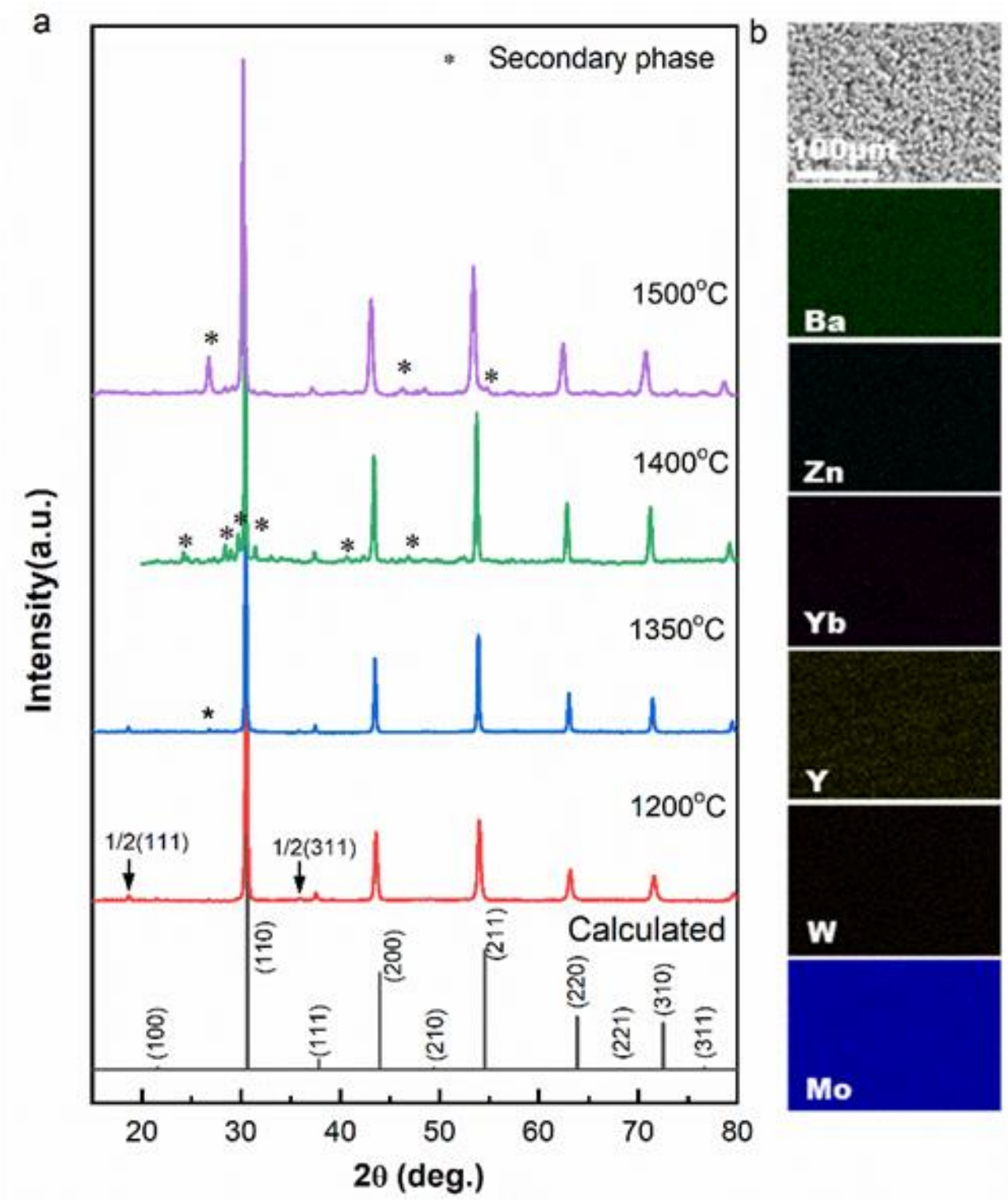

**Figure 2**. (a) XRD patterns of the sample made from composition C2 at various temperatures as labeled. (b) SEM-EDXS elemental maps of the sample made from composition C2 at 1350 $^{o}$C.

**Table 1.** Summary of valence combinations, cation combinations and structural tolerance factor.

| No. | Valence combination (subsystem) | | | | | Cation combination | | | | | $t$ | $\delta(r_B)$ | $\delta(V_B)$ |
|---|---|---|---|---|---|---|---|---|---|---|---|---|---|
| | $B_1$ | $B_2$ | $B_3$ | $B_4$ | $B_5$ | $B_1$ | $B_2$ | $B_3$ | $B_4$ | $B_5$ | | | |
| C1 | +2 | +2 | +4 | +6 | +6 | $Mg^{2+}$ | $Zn^{2+}$ | $Ti^{4+}$ | $W^{6+}$ | $Mo^{6+}$ | 1.04 | 0.10 | 0.45 |
| C2 | +2 | +3 | +3 | +6 | +6 | $Zn^{2+}$ | $Yb^{3+}$ | $Y^{3+}$ | $W^{6+}$ | $Mo^{6+}$ | 0.99 | 0.18 | 0.42 |
| C3 | +2 | +2 | +5 | +5 | +6 | $Mg^{2+}$ | $Zn^{2+}$ | $Nb^{5+}$ | $Ta^{5+}$ | $W^{6+}$ | 1.03 | 0.08 | 0.42 |
| C4 | +2 | +3 | +4 | +5 | +6 | $Zn^{2+}$ | $Yb^{3+}$ | $Ti^{4+}$ | $Nb^{5+}$ | $W^{6+}$ | 1.02 | 0.15 | 0.35 |
| C5 | +2 | +3 | +5 | +5 | +5 | $Zn^{2+}$ | $Yb^{3+}$ | $Nb^{5+}$ | $Ta^{5+}$ | $Sb^{5+}$ | 1.01 | 0.14 | 0.32 |
| C6 | +2 | +4 | +4 | +4 | +6 | $Zn^{2+}$ | $Ti^{4+}$ | $Zr^{4+}$ | $Hf^{4+}$ | $W^{6+}$ | 1.03 | 0.09 | 0.32 |
| C7 | +2 | +4 | +4 | +5 | +5 | $Zn^{2+}$ | $Ti^{4+}$ | $Zr^{4+}$ | $Nb^{5+}$ | $Ta^{5+}$ | 1.03 | 0.08 | 0.27 |
| C8 | +3 | +3 | +3 | +5 | +6 | $Yb^{3+}$ | $Y^{3+}$ | $Ga^{3+}$ | $Nb^{5+}$ | $W^{6+}$ | 1.00 | 0.18 | 0.32 |
| C9 | +3 | +3 | +4 | +4 | +6 | $Yb^{3+}$ | $Y^{3+}$ | $Ti^{4+}$ | $Zr^{4+}$ | $W^{6+}$ | 1.00 | 0.17 | 0.27 |
| C10 | +3 | +3 | +4 | +5 | +5 | $Yb^{3+}$ | $Y^{3+}$ | $Ti^{4+}$ | $Nb^{5+}$ | $Ta^{5+}$ | 1.00 | 0.17 | 0.22 |
| C11 | +3 | +4 | +4 | +4 | +5 | $Yb^{3+}$ | $Zr^{4+}$ | $Ti^{4+}$ | $Sn^{4+}$ | $Nb^{5+}$ | 1.01 | 0.13 | 0.16 |
| C12 | +4 | +4 | +4 | +4 | +4 | $Zr^{4+}$ | $Ti^{4+}$ | $Ce^{4+}$ | $Hf^{4+}$ | $Sn^{4+}$ | 1.00 | 0.12 | 0.00 |

**Table 2.** Phase composition and lattice parameters of the synthesized samples.

| No | Composition | Second phase? | | | | | Lattice (Å) |
|---|---|---|---|---|---|---|---|
| | | 1200 ℃ | 1350 ℃ | 1400 ℃ | 1500 ℃ | 1600 ℃ | |
| C1 | $Ba(Mg_{0.2}Zn_{0.2}Ti_{0.2}W_{0.2}Mo_{0.2})O_3$ | Major | Major | Major | - | - | - |
| C2 | $Ba(Zn_{0.2}Yb_{0.2}Y_{0.2}W_{0.2}Mo_{0.2})O_3$ | NO/PO | Minor/PO | Major | Major | - | 4.1790 |
| C3 | $Ba(Mg_{0.2}Zn_{0.2}Nb_{0.2}Ta_{0.2}W_{0.2})O_3$ | NO/PO | NO/PO | - | Major | NO/OP | 4.0862 |
| C4 | $Ba(Zn_{0.2}Yb_{0.2}Ti_{0.2}Nb_{0.2}W_{0.2})O_3$ | No | Minor | - | Minor | Major | 4.1296 |
| C5 | $Ba(Zn_{0.2}Yb_{0.2}Nb_{0.2}Ta_{0.2}Sb_{0.2})O_3$ | Major | Major | - | Major | Major | - |
| C6 | $Ba(Zn_{0.2}Ti_{0.2}Zr_{0.2}Hf_{0.2}W_{0.2})O_3$ | Major | Major | - | No | Major | 4.1104 |
| C7 | $Ba(Zn_{0.2}Ti_{0.2}Zr_{0.2}Nb_{0.2}Ta_{0.2})O_3$ | Minor | No | - | Minor | Minor | 4.1089 |
| C8 | $Ba(Yb_{0.2}Y_{0.2}Ga_{0.2}Nb_{0.2}W_{0.2})O_3$ | No | Major | - | No | Major | 4.1954 |
| C9 | $Ba(Yb_{0.2}Y_{0.2}Ti_{0.2}Zr_{0.2}W_{0.2})O_3$ | No | No | - | No | No | 4.1877 |
| C10 | $Ba(Yb_{0.2}Y_{0.2}Ti_{0.2}Nb_{0.2}Ta_{0.2})O_3$ | No | No | - | Minor | Major | 4.1777 |
| C11 | $Ba(Yb_{0.2}Ti_{0.2}Zr_{0.2}Sn_{0.2}Nb_{0.2})O_3$ | - | No | - | Major | Major | 4.1512 |
| C12 | $Ba(Ti_{0.2}Zr_{0.2}Hf_{0.2}Sn_{0.2}Ce_{0.2})O_3$ | Major | - | No | Minor | Minor | 4.1551 |

The sample with secondary phase(s) that could not be identified by XRD is labeled as “NO”. If secondary phase(s) could be identified by XRD, the sample with the intensity of the strongest XRD peak of secondary phases < 4% of that of the perovskite (110) peak is labeled as “Minor”; and the sample with the intensity of the strongest XRD peak of secondary phases > 4% of that of the perovskite (110) peak is labeled as “Major”. The sample containing partially ordering structure is labeled as PO.

**Table 3**. Electrical properties of the high-entropy perovskite ceramics.

| No | Composition | Sintering temperature (ºC) | Dielectric constant ε (@1kHz) | Dielectric loss tanδ (1kHz) | Activation energy, $E_a$ (eV) | AC Conductivity at 666ºC and 1 kHz (S/m) |
|---|---|---|---|---|---|---|
| C2 | $Ba(Zn_{0.2}Yb_{0.2}Y_{0.2}W_{0.2}Mo_{0.2})O_3$ | 1350 | 94.1 | 1.47 | 1.20 | $2.77*10^{-5}$ |
| C3 | $Ba(Mg_{0.2}Zn_{0.2}Nb_{0.2}Ta_{0.2}W_{0.2})O_3$ | 1350 | 149.2 | 1.60 | 1.16 | $2.49*10^{-6}$ |
| C4 | $Ba(Zn_{0.2}Yb_{0.2}Ti_{0.2}Nb_{0.2}W_{0.2})O_3$ | 1350 | 112.8 | 0.97 | 0.69 | $2.41*10^{-5}$ |
| C6 | $Ba(Zn_{0.2}Ti_{0.2}Zr_{0.2}Hf_{0.2}W_{0.2})O_3$ | 1500 | 526.0 | 2.63 | 0.53 | $1.60*10^{-5}$ |
| C7 | $Ba(Zn_{0.2}Ti_{0.2}Zr_{0.2}Nb_{0.2}Ta_{0.2})O_3$ | 1500 | 48.2 | 0.35 | 1.20 | $2.07*10^{-7}$ |
| C8 | $Ba(Yb_{0.2}Y_{0.2}Ga_{0.2}Nb_{0.2}W_{0.2})O_3$ | 1200 | 25.4 | 0.03 | 1.21 | $2.92*10^{-6}$ |
| C9 | $Ba(Yb_{0.2}Y_{0.2}Ti_{0.2}Zr_{0.2}W_{0.2})O_3$ | 1500 | 64.7 | 0.29 | 1.00 | $1.63*10^{-5}$ |
| C10 | $Ba(Yb_{0.2}Y_{0.2}Ti_{0.2}Nb_{0.2}Ta_{0.2})O_3$ | 1500 | 46.5 | 0.003 | 0.98 | $4.33*10^{-6}$ |
| C11 | $Ba(Yb_{0.2}Ti_{0.2}Zr_{0.2}Sn_{0.2}Nb_{0.2})O_3$ | 1350 | 52.8 | 0.16 | 1.15 | $3.81*10^{-6}$ |
| C12 | $Ba(Ti_{0.2}Zr_{0.2}Hf_{0.2}Sn_{0.2}Ce_{0.2})O_3$ | 1400 | 277.3 | 0.58 | 0.67 | $3.87*10^{-6}$ |

# Supporting Information

## High-Entropy Oxides Based on Valence Combinations: Design and Practice

*Lei Tang, Zemin Li, Kepi Chen*, Cuiwei Li, Xiaowen Zhang, and Linan An**

**Table S1**: 8 independent valence combinations in an $A(4B_{0.25})O_3$ perovskite system.

| No | $B_1$ | $B_2$ | $B_3$ | $B_4$ |
|---|---|---|---|---|
| #1 | +2 | +2 | +6 | +6 |
| #2 | +2 | +3 | +5 | +6 |
| #3 | +2 | +4 | +4 | +6 |
| #4 | +2 | +4 | +5 | +5 |
| #5 | +3 | +3 | +4 | +6 |
| #6 | +3 | +3 | +5 | +5 |
| #7 | +3 | +4 | +4 | +5 |
| #8 | +4 | +4 | +4 | +4 |

**Table S2**: 18 independent valence combinations for an $A(6B_{1/6})O_3$ perovskite system.

| No | $B_1$ | $B_2$ | $B_3$ | $B_4$ | $B_5$ | $B_6$ |
|---|---|---|---|---|---|---|
| #1 | +2 | +2 | +2 | +6 | +6 | +6 |
| #2 | +2 | +2 | +3 | +5 | +6 | +6 |
| #3 | +2 | +2 | +4 | +4 | +6 | +6 |
| #4 | +2 | +2 | +4 | +5 | +5 | +6 |
| #5 | +2 | +2 | +5 | +5 | +5 | +5 |
| #6 | +2 | +3 | +3 | +4 | +6 | +6 |
| #7 | +2 | +3 | +3 | +5 | +5 | +6 |
| #8 | +2 | +3 | +4 | +4 | +5 | +6 |
| #9 | +2 | +3 | +4 | +5 | +5 | +5 |
| #10 | +2 | +4 | +4 | +4 | +4 | +6 |
| #11 | +2 | +4 | +4 | +4 | +5 | +5 |
| #12 | +3 | +3 | +3 | +3 | +6 | +6 |
| #13 | +3 | +3 | +3 | +4 | +5 | +6 |
| #14 | +3 | +3 | +3 | +5 | +5 | +5 |
| #15 | +3 | +3 | +4 | +4 | +4 | +6 |
| #16 | +3 | +3 | +4 | +4 | +5 | +5 |
| #17 | +3 | +4 | +4 | +4 | +4 | +5 |
| #18 | +4 | +4 | +4 | +4 | +4 | +4 |

**Table S3**: The coordination number and ionic radius of the ions used in this study.

| Ion | Coordination number | Radius（Å） |
|---|---|---|
| $Ba^{2+}$ | 12 | 1. 61 |
| $O^{2-}$ | 6 | 1.40 |
| $Mg^{2+}$ | 6 | 0.72 |
| $Zn^{2+}$ | 6 | 0.74 |
| $Ga^{3+}$ | 6 | 0.62 |
| $Y^{3+}$ | 6 | 0.90 |
| $Yb^{3+}$ | 6 | 0.868 |
| $Ce^{4+}$ | 6 | 0.87 |
| $Hf^{4+}$ | 6 | 0.71 |
| $Sn^{4+}$ | 6 | 0.69 |
| $Ti^{4+}$ | 6 | 0.605 |
| $Zr^{4+}$ | 6 | 0.72 |
| $Nb^{5+}$ | 6 | 0.64 |
| $Sb^{5+}$ | 6 | 0.60 |
| $Ta^{5+}$ | 6 | 0.64 |
| $Mo^{6+}$ | 6 | 0.59 |
| $W^{6+}$ | 6 | 0.60 |

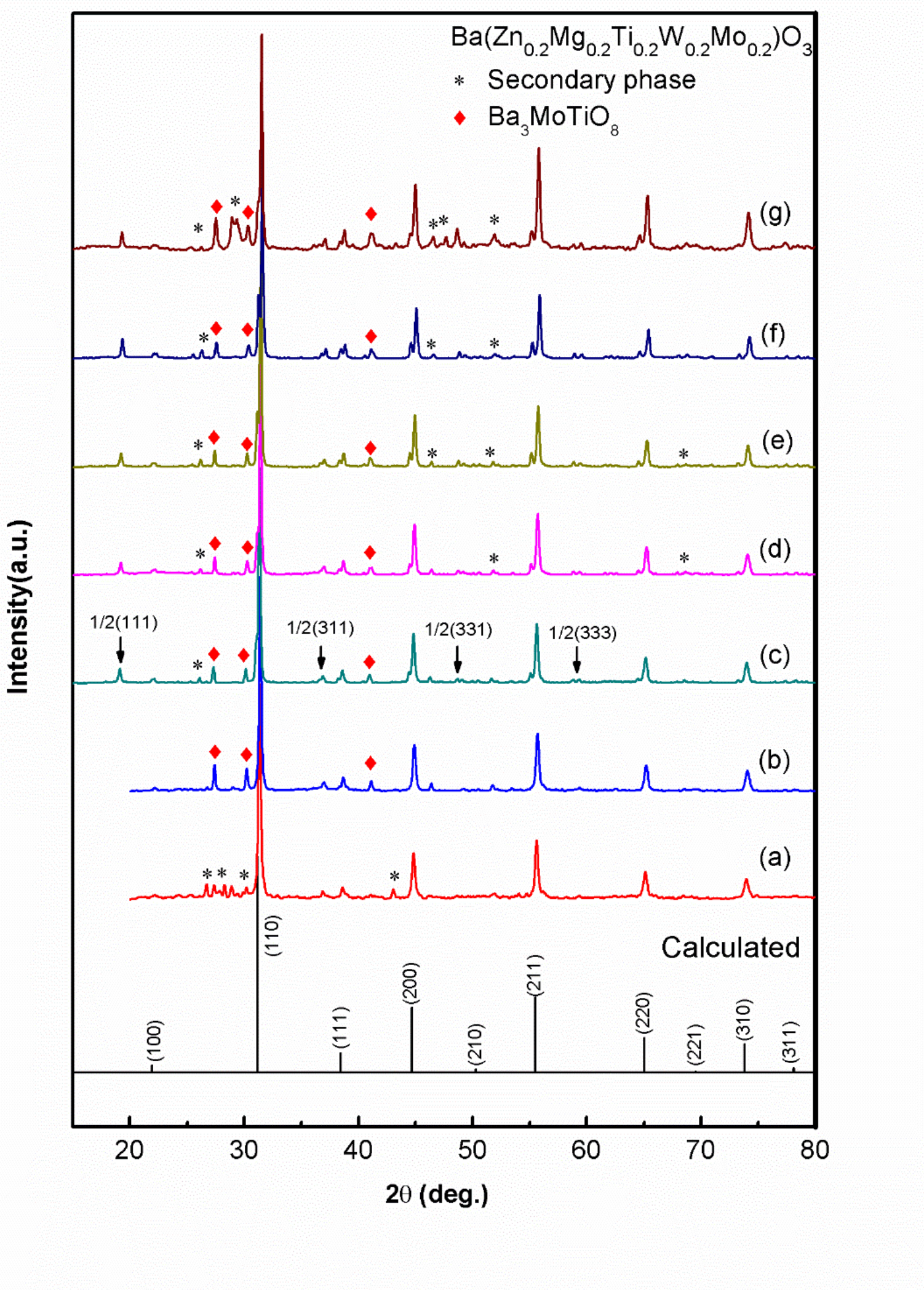

**Figure S1**. XRD patterns experimentally measured on the samples made from composition C1 at different temperatures: (a) powder calcined at 800 °C for 6 hrs, (b) powder calcined at 1000 °C for 6 hrs, (c) powder calcined at 600 °C for 6 hrs and then at 1100 °C for 6 hrs, (d) powder calcined at 700 °C for 6 hrs and then at 1100 °C for 6 hrs, (e) powder calcined at 1200 °C for 6 hours, (f) powder calcined at 1350 °C for 6 hrs, and (g) ceramic sample sintered at 1400 °C for 6 hrs. The XRD pattern calculated using the rule of mixture by assuming equal and random distribution of five substituted cations is also included. The results reveal that the composition cannot form a single-phase compound.

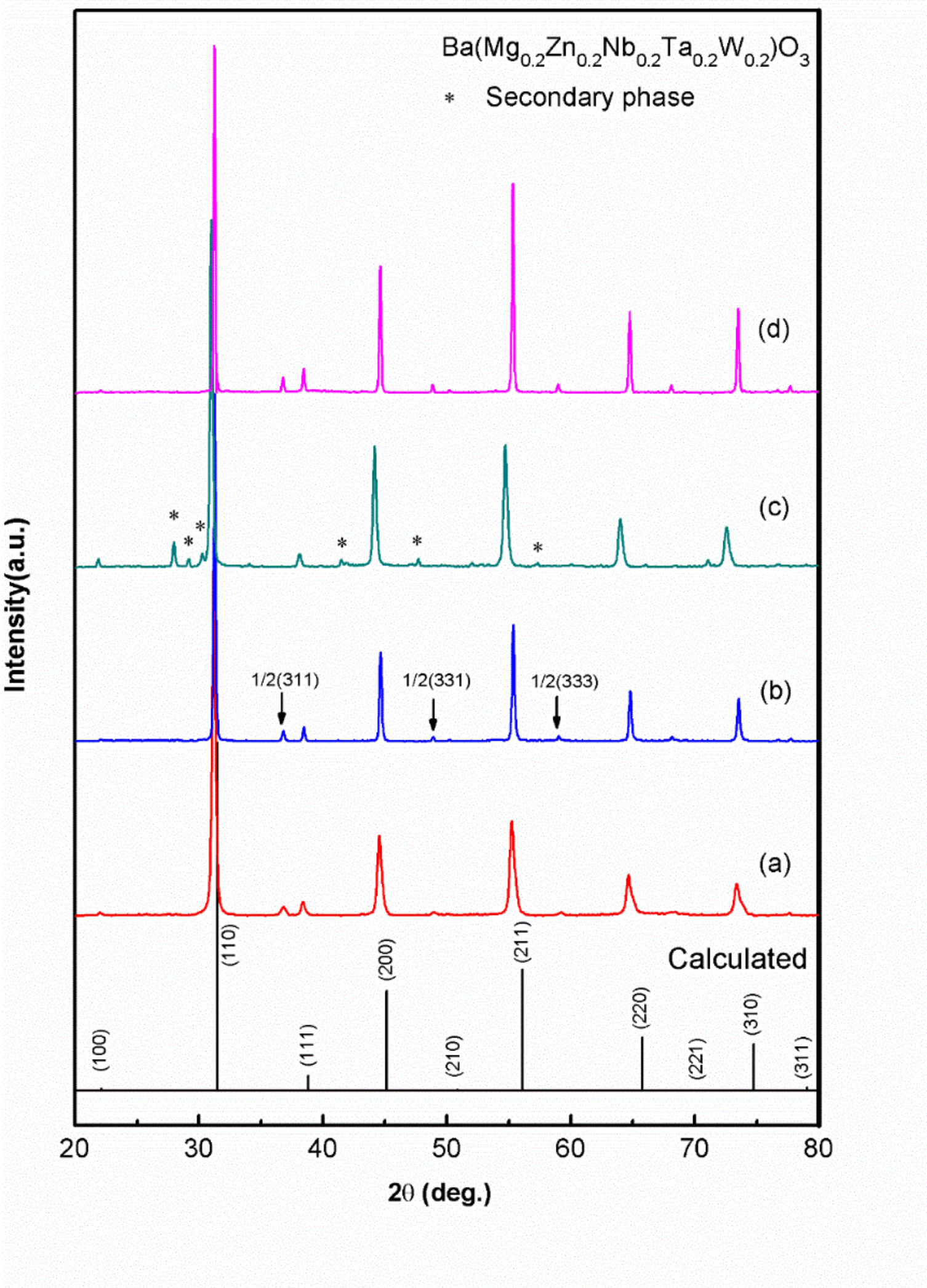

**Figure S2**. XRD patterns experimentally measured on the samples made of composition C3 at different temperatures: (a) powder calcined at 1200 °C for 12 hrs, (b) powder calcined at 1350 °C for 12 hrs, (c) ceramic sample sintered at 1500 °C for 6 hrs, and (d) ceramic sample sintered at 1600 °C for 4 hrs. The XRD pattern calculated using the rule of mixture by assuming equal and random distribution of five substituted cations is also included. The results reveal that the composition can form a single-phase perovskite with the chemical ordering of the B-site.

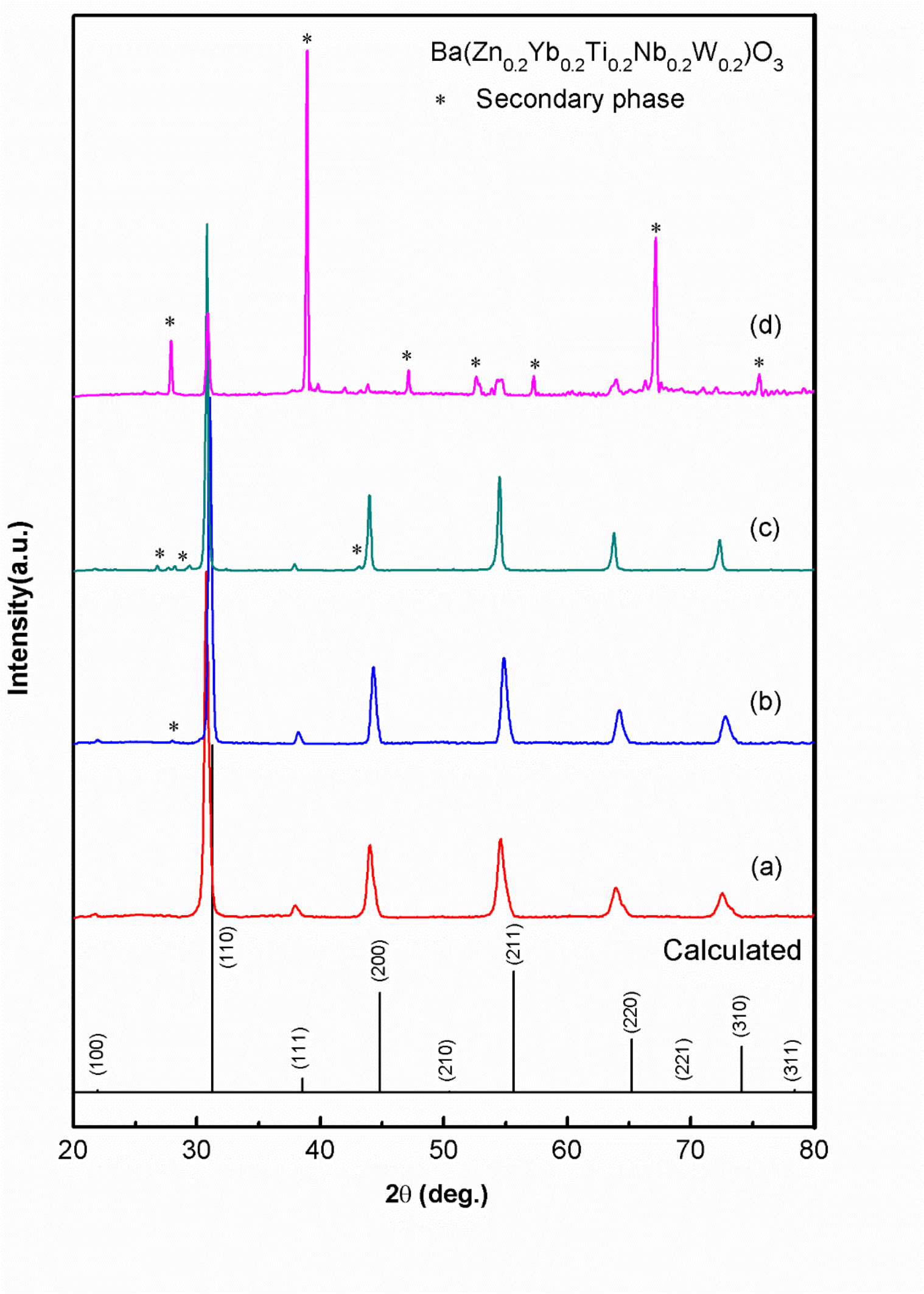

**Figure S3**. XRD patterns experimentally measured on the samples made from composition C4 at different temperatures: (a) powder calcined at 1200 °C for 12 hrs, (b) powder calcined at 1350 °C for 12 hrs, (c) ceramic sample sintered at 1500 °C for 6 hrs, and (d) ceramic sample sintered at 1600 °C for 4 hrs. The XRD pattern calculated using the rule of mixture by assuming equal and random distribution of five substituted cations is also included. The results reveal that the composite can form a single-phase cubic perovskite.

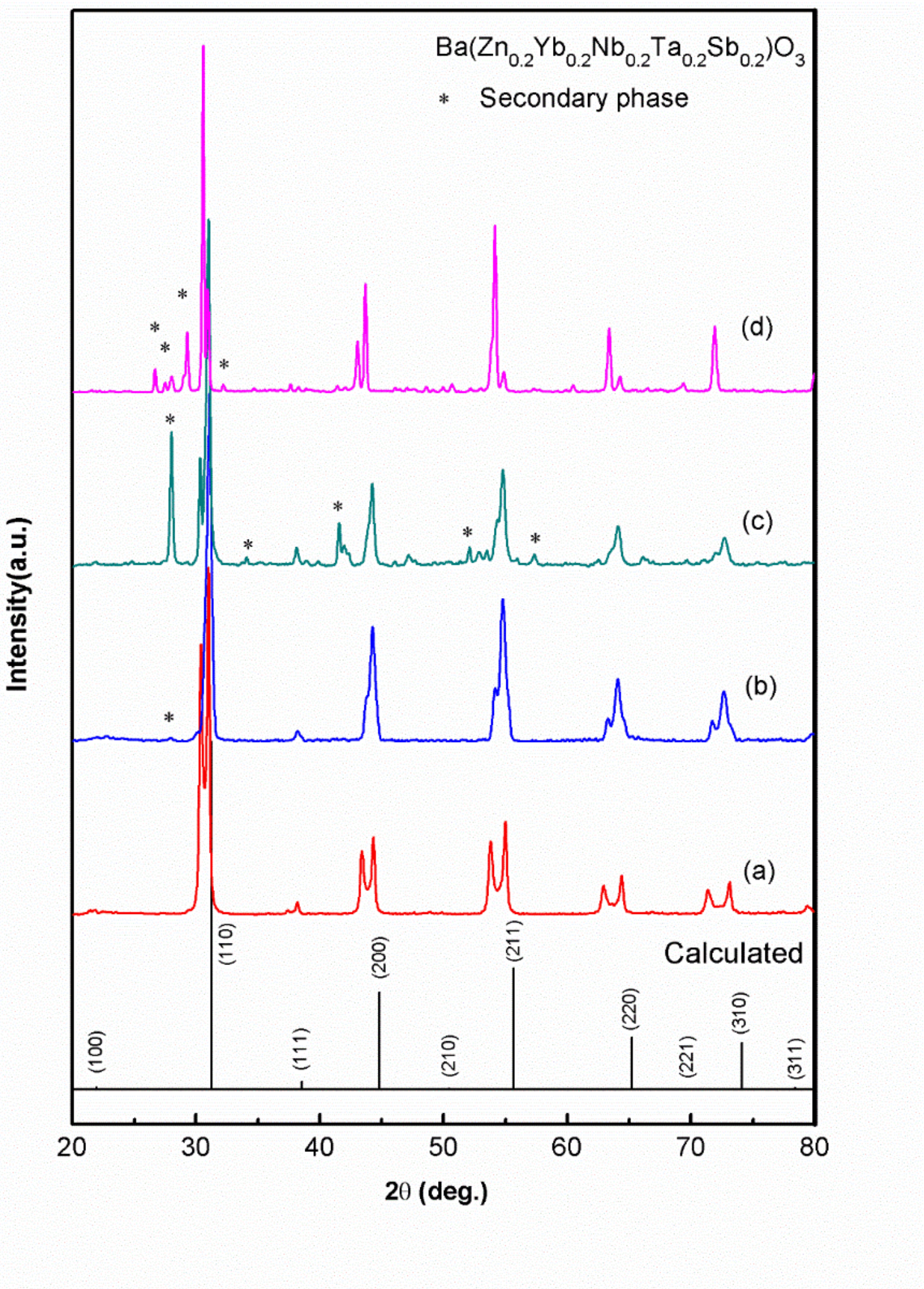

**Figure S4.** XRD patterns experimentally measured on the samples made from composition C5 at different temperatures: (a) powder calcined at 1200 °C for 12 hrs, (b) powder calcined at 1350 °C for 12 hrs, (c) ceramic sample sintered at 1500 °C for 6 hrs, and (d) ceramic sample sintered at 1600 °C for 4 hrs. The XRD pattern calculated using the rule of mixture by assuming equal and random distribution of five substituted cations is also included. The results reveal that the composition cannot form a single-phase compound.

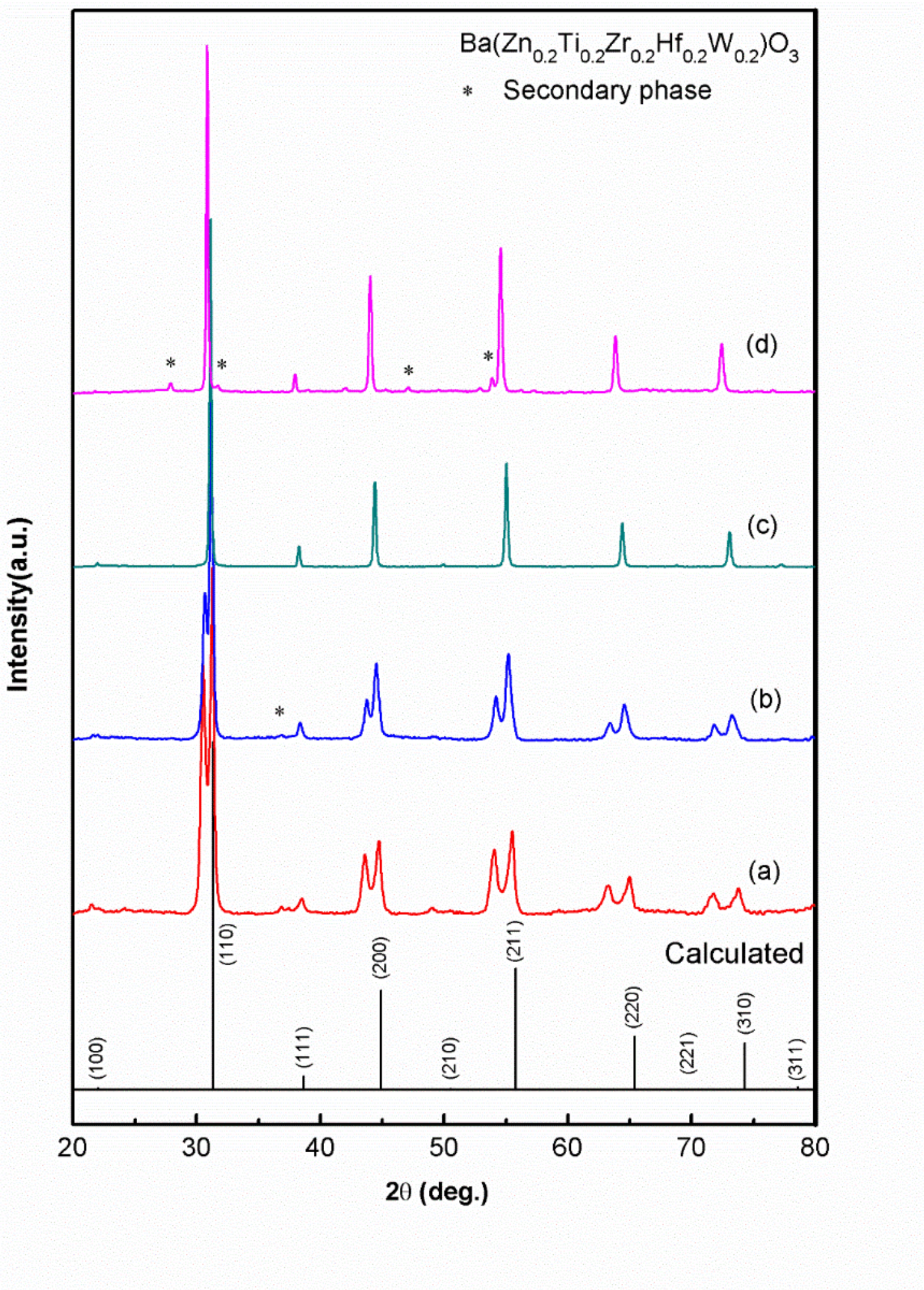

**Figure S5**. XRD patterns experimentally measured on the samples made from composition C6 at different temperatures: (a) powder calcined at 1200 °C for 12 hrs, (b) powder calcined at 1350 °C for 12 hrs, (c) ceramic sample sintered at 1500 °C for 6 hrs, and (d) ceramic sample sintered at 1600 °C for 4 hrs. The XRD pattern calculated using the rule of mixture by assuming equal and random distribution of five substituted cations is also included. The results reveal that the composition can form a single-phase cubic perovskite.

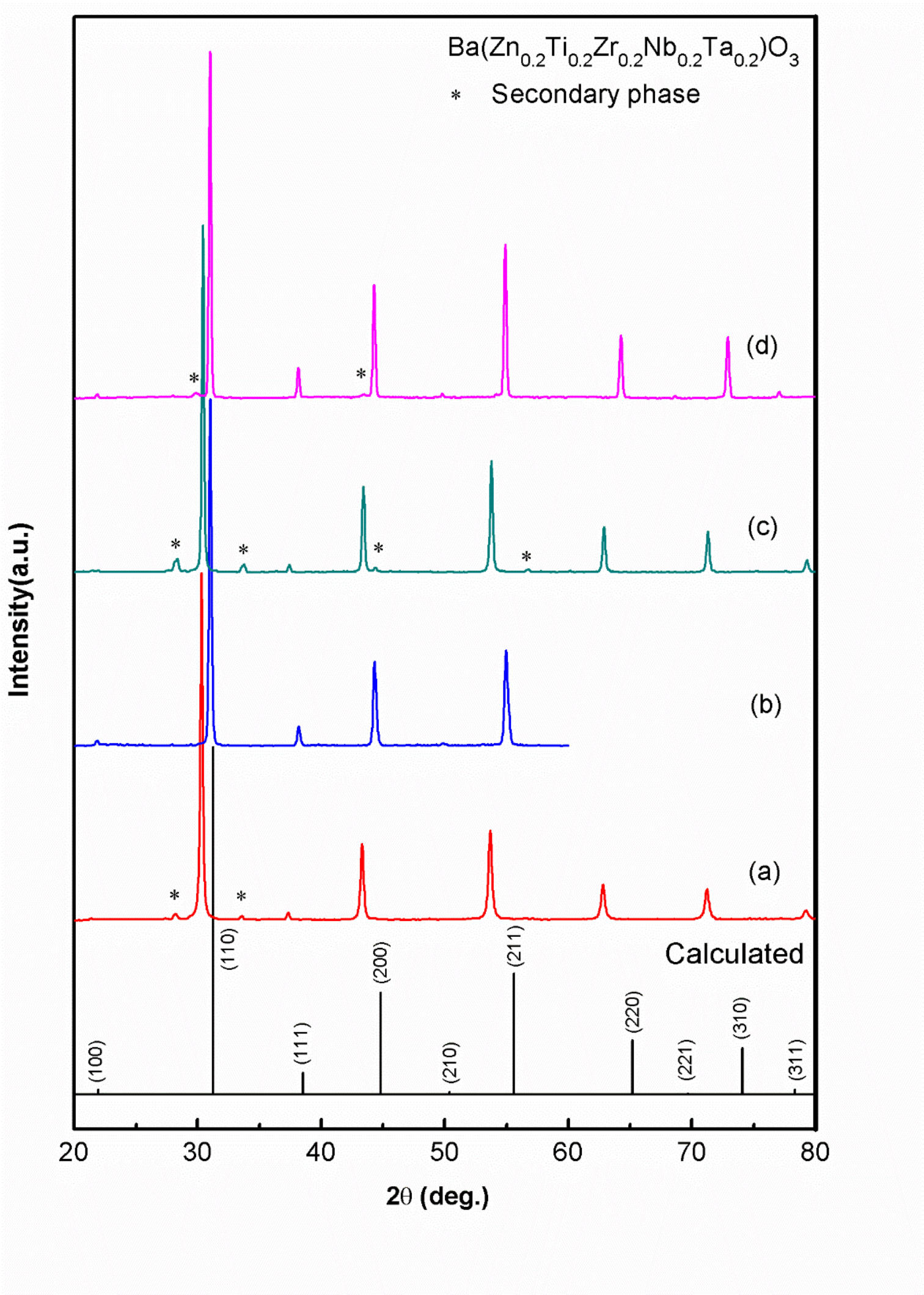

**Figure S6**. XRD patterns experimentally measured on the samples made of composition C7 at different temperatures: (a) powder calcined at 1200 °C for 12 hrs, (b) powder calcined at 1350 °C for 12 hrs, (c) ceramic sample sintered at 1500 °C for 6 hrs, and (d) ceramic sample sintered at 1600 °C for 4 hrs. The XRD pattern calculated using the rule of mixture by assuming equal and random distribution of five substituted cations is also included. The results reveal that the composition can form a single-phase cubic perovskite.

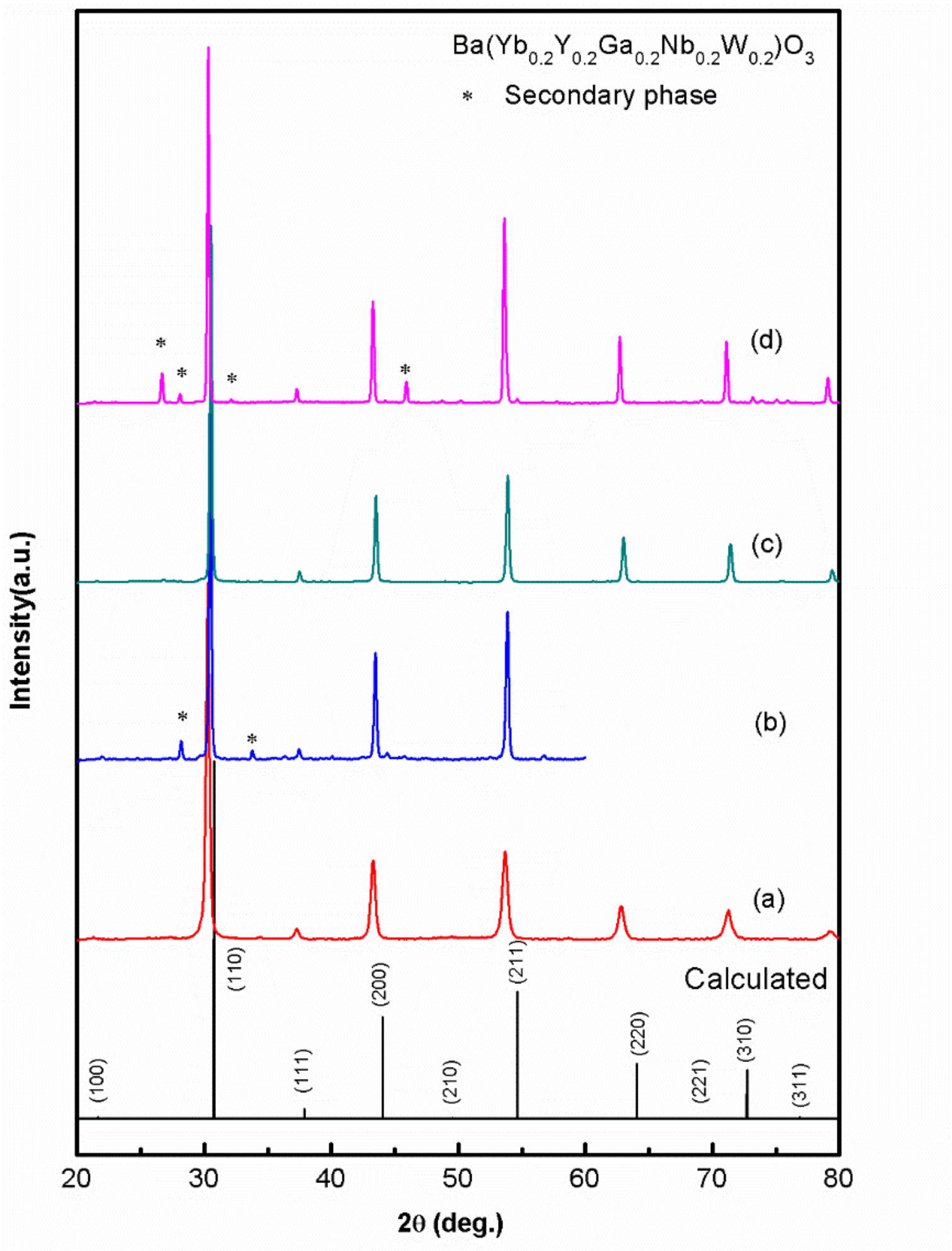

**Figure S7**. XRD patterns experimentally measured on the samples made from composition C8 at different temperatures: (a) powder calcined at 1200 °C for 12 hrs, (b) powder calcined at 1350 °C for 12 hrs, (c) ceramic sample sintered at 1500 °C for 6 hrs, and (d) ceramic sample sintered at 1600 °C for 4 hrs. The XRD pattern calculated using the rule of mixture by assuming equal and random distribution of five substituted cations is also included. The results reveal that the composition can form a single-phase cubic perovskite.

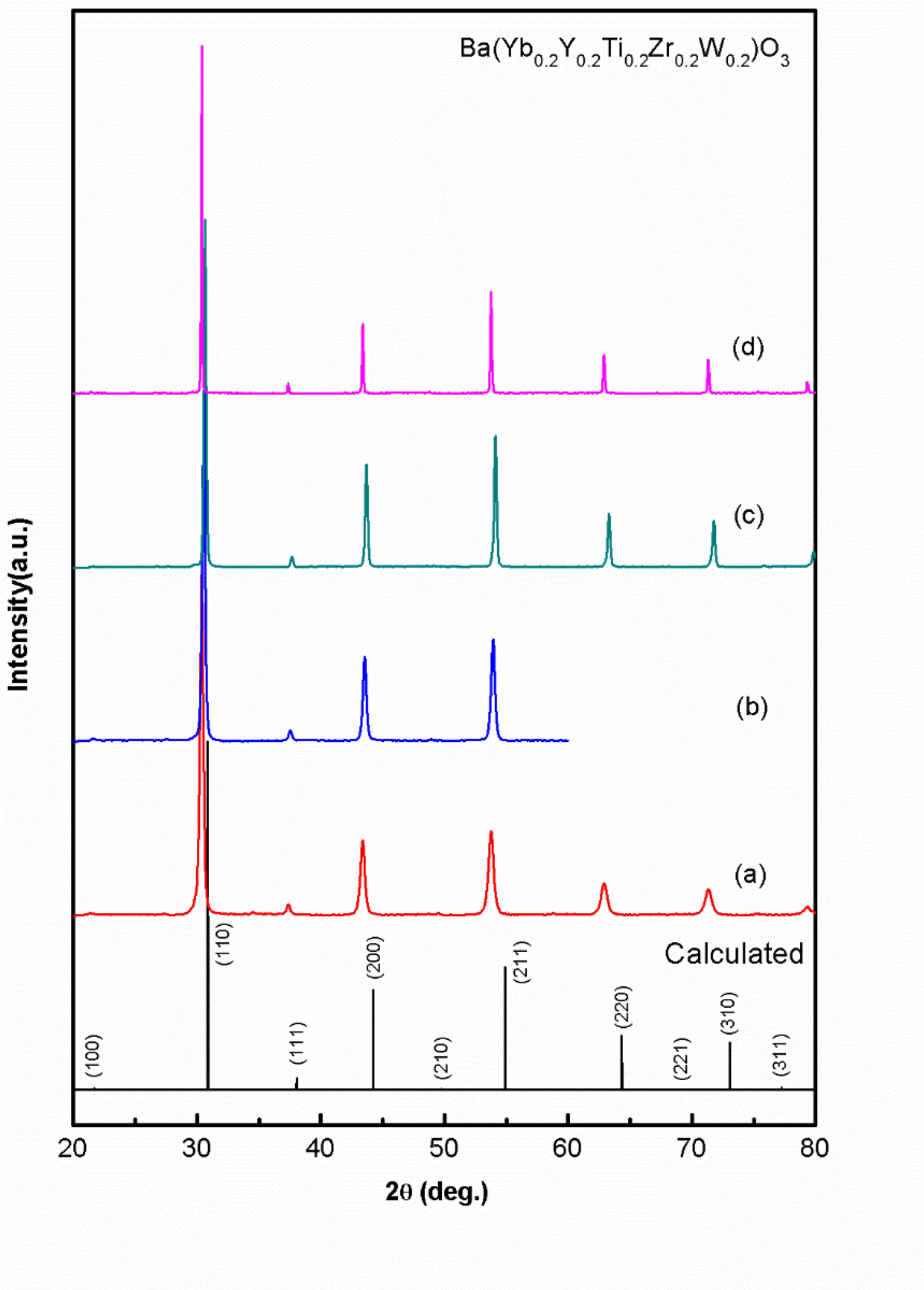

**Figure S8**. XRD patterns departmentally measured on the samples made from composition C9 at different temperatures: (a) powder calcined at 1200 °C for 12 hrs, (b) powder calcined at 1350 °C for 12 hrs, (c) ceramic sample sintered at 1500 °C for 6 hrs, and (d) ceramic sample sintered at 1600 °C for 4 hrs. The XRD pattern calculated using the rule of mixture by assuming equal and random distribution of five substituted cations is also included. The results reveal that the composition can form a single-phase cubic perovskite.

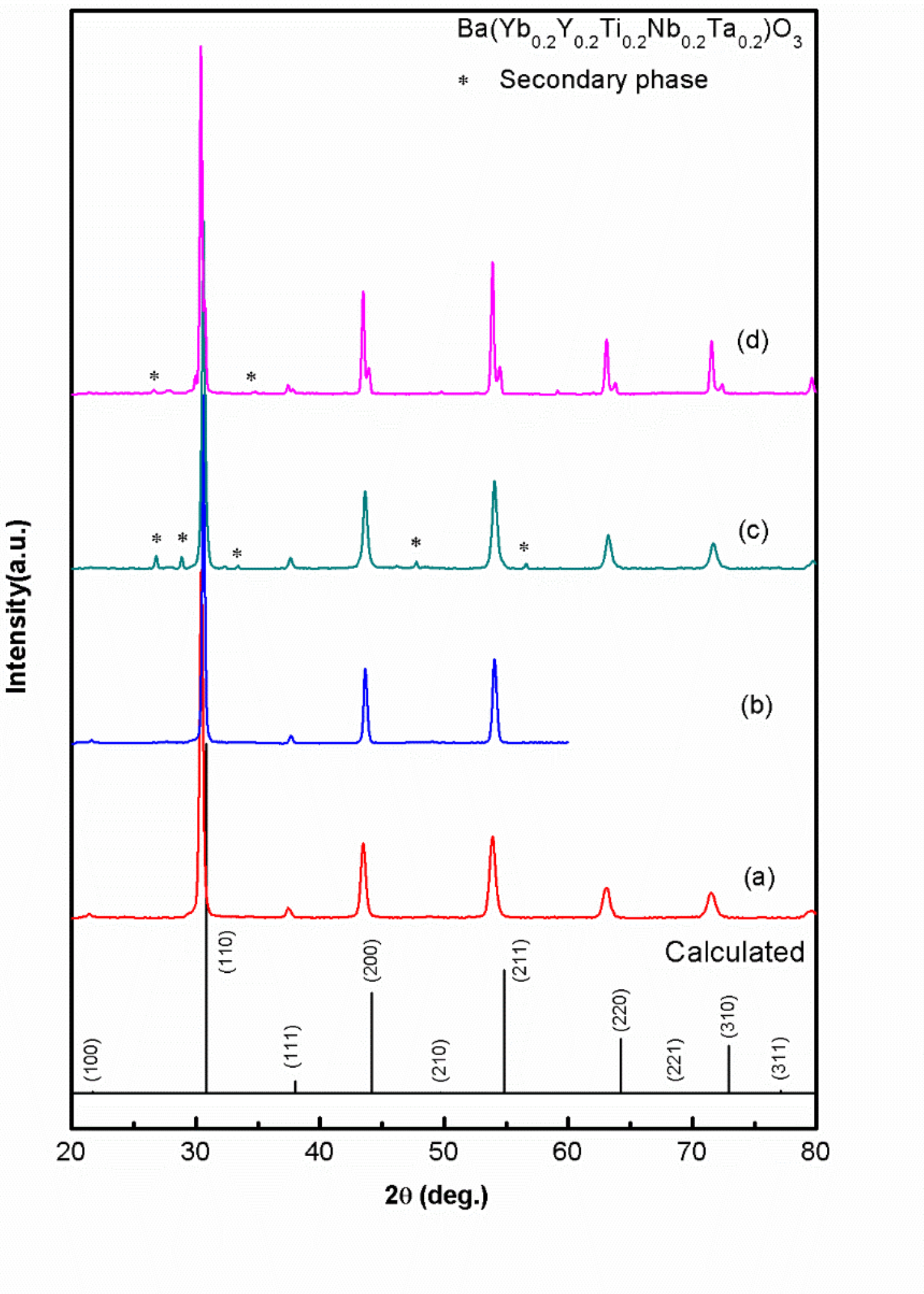

**Figure S9**. XRD patterns experimentally measured on the samples made from composition C10 at different temperatures: (a) powder calcined at 1200 °C for 12 hrs, (b) powder calcined at 1350 °C for 12 hrs, (c) ceramic sample sintered at 1500 °C for 6 hrs, and (d) ceramic sample sintered at 1600 °C for 4 hrs. The XRD pattern calculated using the rule of mixture by assuming equal and random distribution of five substituted cations is also included. The results reveal that the composition can form a single-phase cubic perovskite.

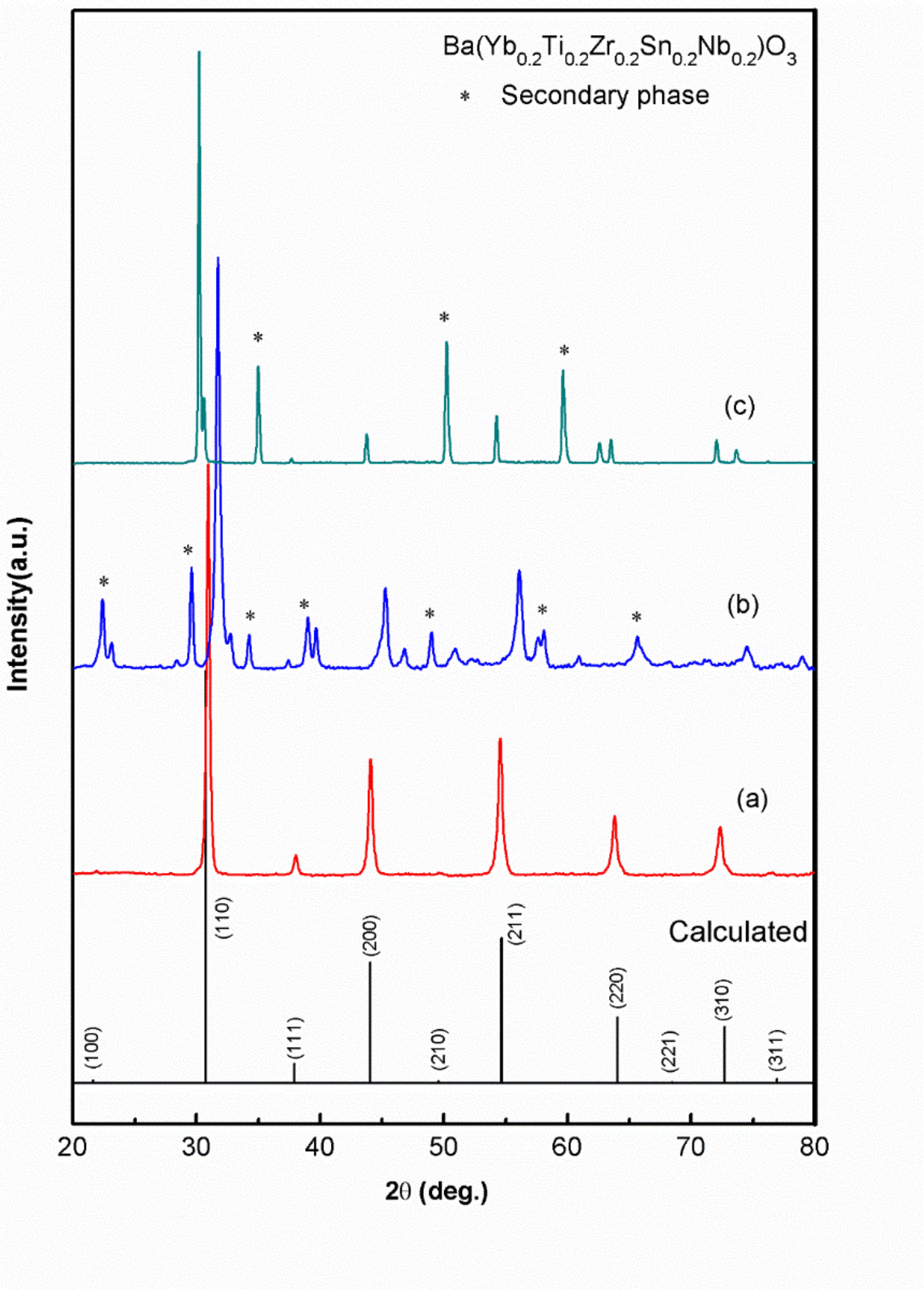

**Figure S10**. XRD patterns experimentally measured on the samples made from composition C11 at different temperatures: (a) powder calcined at 1350 °C for 12 hrs, (b) ceramic sample sintered at 1500 °C for 6 hrs, and (c) ceramic sample sintered at 1600 °C for 4 hrs. The XRD pattern calculated using the rule of mixture by assuming equal and random distribution of five substituted cations is also included. The results reveal that the composition can form a single-phase cubic perovskite.